# Software Design Pattern Model and Data Structure Algorithm Abilities on Microservices Architecture Design in High-tech Enterprises.

Jun Cui[1, a, *]

1 Solbridge International School of Business, Ph.D., Daejeon, 34613, Republic of Korea.

a jcui228@student.solbridge.ac.kr

* Correspondence: Jun Cui, jcui228@student.solbridge.ac.kr.

*Abstract*—This study investigates the impact of software design model capabilities and data structure algorithm abilities on microservices architecture design within enterprises. Utilizing a qualitative methodology, the research involved in-depth interviews with software architects and developers who possess extensive experience in microservices implementation. The findings reveal that organizations emphasizing robust design models and efficient algorithms achieve superior scalability, performance, and flexibility in their microservices architecture. Notably, participants highlighted that a strong foundation in these areas facilitates better service decomposition, optimizes data processing, and enhances system responsiveness. Despite these insights, gaps remain regarding the integration of emerging technologies and the evolving nature of software design practices. This paper contributes to the existing literature by underscoring the critical role of these competencies in fostering effective microservices architectures and suggests avenues for future research to address identified gaps.

*Keywords—Microservices architecture, software design pattern models, data structures, algorithms, enterprise software development, qualitative research, scalability, performance.*

## I. INTRODUCTION

The evolution of software architecture has witnessed a significant shift from monolithic systems to microservices architectures, driven by the need for greater flexibility, scalability, and rapid deployment. Microservices allow organizations to decompose complex applications into smaller, independently deployable services that can be developed, updated, and scaled without affecting the entire system. This paradigm shift is particularly pertinent in today's fast-paced business environment, where responsiveness to market changes is crucial [1]. However, the successful implementation of microservices relies heavily on the foundational competencies of software design models and data structure algorithms. These capabilities enable developers to create systems that are not only functional but also maintainable and efficient. Understanding how these competencies influence microservices design is vital for enterprises aiming to leverage this architectural style to enhance their operational efficiency and competitiveness [2].

### Research Problems and Gaps

Despite the growing adoption of microservices, organizations often face challenges related to service orchestration, data management, and system integration. Many enterprises struggle with ensuring that their microservices communicate effectively and that data flows seamlessly across services [3]. Furthermore, the lack of a clear understanding of design principles and the optimal use of data structures can lead to inefficiencies, increased latency, and difficulties in scaling applications. Current literature lacks comprehensive insights into the specific ways in which software design models and data structure algorithms impact the effectiveness of microservices architectures. Additionally, there is limited exploration of how emerging technologies, such as cloud computing and containerization, interact with these foundational competencies, creating a critical gap that this research aims to address [1-3].

### Research Questions

To navigate the identified gaps, this study is guided by several research questions aimed at understanding the relationship between design capabilities and microservices architecture.

RQ1. First, how do software design model capabilities influence the scalability and performance of microservices?

RQ2. Second, what role do data structure algorithms play in optimizing microservices communication and data handling?

RQ3. Third, how can organizations enhance their design and algorithm capabilities to better leverage microservices architecture?

These questions will guide the exploration of the interplay between theoretical foundations and practical applications, providing a clearer understanding of how these competencies affect microservices design and implementation. By addressing these queries, the study seeks to illuminate the pathways through which enterprises can achieve greater efficiency and agility in their software development practices.

### Motivations

The motivation for this research stems from the increasing importance of microservices in modern software development and the recognition that foundational competencies in design models and algorithms are critical to their success [4]. As enterprises strive to remain competitive in a dynamic marketplace, the ability to effectively implement microservices can be a game-changer. This study aims to provide insights that will help organizations understand the value of investing in software design and algorithm capabilities. By highlighting the correlation between these competencies and successful microservices architecture, the research seeks to inform strategic decisions regarding resource allocation, training, and development. Ultimately, the goal is to empower organizations to harness the full potential of microservices, thereby enhancing their operational agility and responsiveness [3-6].

**Paper Structure**

This paper is structured to provide a comprehensive exploration of the impact of software design model and data structure algorithm capabilities on microservices architecture design. Following this introduction, the literature review will present relevant theories and previous research, establishing a theoretical framework for the study. The methodology section will detail the qualitative approach employed, including participant selection and data collection methods. The results section will summarize key findings from the interviews, highlighting the experiences and perspectives of industry professionals. Finally, the discussion will interpret these findings in the context of existing literature, addressing the implications for practice and identifying avenues for future research. This structured approach ensures a thorough examination of the topic, ultimately contributing valuable insights to the field of enterprise software development.

## II. LITERATURE REVIEW

**Microservices Architecture**

Microservices architecture has become a pivotal response to the challenges posed by monolithic application structures, where an entire system is built as a single, indivisible unit. This traditional approach often leads to complications in deployment, scaling, and maintaining applications as they grow. Microservices architecture, in contrast, breaks down applications into smaller, autonomous services that communicate over well-defined APIs. Each microservice encapsulates a specific business function, allowing teams to develop, deploy, and scale these services independently. This modular approach offers several advantages, including improved scalability, enhanced fault isolation, and accelerated development cycles [5-6].

One of the fundamental principles of microservices is service autonomy. Each microservice operates independently, which means that developers can update, deploy, or scale services without impacting the entire application. This decoupling allows organizations to implement continuous integration and continuous delivery (CI/CD) practices more effectively. By adopting CI/CD, teams can automate testing and deployment, leading to faster release cycles and more reliable software delivery. Furthermore, service autonomy fosters innovation, as different teams can experiment with new technologies and approaches without being constrained by a monolithic codebase [7].

Another critical aspect of microservices architecture is decentralized data management. In contrast to monolithic systems, where a single database is often shared, microservices advocate for each service to manage its own data. This decentralization enhances data integrity and reduces the risk of data contention, as services are less likely to interfere with each other's data operations. Additionally, it allows for more flexible data storage solutions tailored to the specific needs of each microservice. For instance, one service may utilize a relational database for structured data, while another might employ a NoSQL database for unstructured data. This flexibility supports a more efficient allocation of resources and enables organizations to adopt the best technologies for their unique requirements [5].

Lastly, the principle of continuous delivery is integral to microservices architecture. Continuous delivery ensures that code changes are automatically tested and deployed to production environments, promoting a culture of frequent, incremental improvements. This practice not only enhances software quality but also allows organizations to respond swiftly to market demands and user feedback. By fostering a responsive development environment, enterprises can maintain a competitive edge in today's fast-paced technological landscape. In summary, microservices architecture presents a robust solution to the limitations of monolithic systems by emphasizing service autonomy, decentralized data management, and continuous delivery, thereby enabling organizations to build scalable, maintainable, and innovative applications [5-9].

**Software Design pattern Models**

Software design models serve as essential blueprints in the development process, guiding architects and developers in creating systems that are both scalable and maintainable. By providing structured approaches to design, these models facilitate effective communication among team members and promote best practices in software engineering. Among the most prominent design models pertinent to microservices are Domain-Driven Design (DDD) and Model-View-Controller (MVC). Each of these models plays a unique role in shaping the architecture of microservices, influencing how services are structured and how they interact with one another [2-3].

Domain-Driven Design (DDD) emphasizes a model-driven approach to software development, where the focus is on the business domain rather than the technology. In DDD, the application is divided into bounded contexts, each representing a specific subdomain of the overall business model. This division enables teams to develop services that are closely aligned with business needs, ensuring that each microservice encapsulates a specific aspect of the domain. DDD promotes collaboration between technical and non-technical stakeholders, fostering a shared understanding of the business requirements and leading to more relevant and effective solutions. Additionally, DDD encourages the use of rich domain models, which can lead to more expressive and maintainable code [6-7].

On the other hand, the Model-View-Controller (MVC) design pattern provides a framework for separating concerns within an application. By dividing the application into three interconnected components—Model, View, and Controller—MVC allows for greater modularity and ease of maintenance. In the context of microservices, MVC can be particularly beneficial as it enables developers to design services that handle specific aspects of the application. For example, one microservice might manage the data model and business logic (Model), while another might handle user interfaces and presentation logic (View). The Controller acts as an intermediary, managing the flow of data between the Model and the View. This separation of concerns simplifies development and testing processes, as each component can be developed and maintained independently [7].

In summary, software design models such as Domain-Driven Design and Model-View-Controller provide critical frameworks for developing microservices architectures. They offer structured approaches to designing software systems, fostering collaboration, enhancing modularity, and promoting best practices. By leveraging these design models, organizations can build scalable and maintainable microservices that align with their business objectives, ultimately improving their software development processes and enhancing overall system performance [5-9].

**Data Structures and Algorithms**

Data structures and algorithms are foundational elements of computer science that significantly impact the performance and efficiency of microservices architectures. As microservices involve the processing and exchange of data across different services, the choice of appropriate data structures and algorithms becomes paramount in ensuring that services operate smoothly and responsively. This section explores common data structures, such as arrays, linked lists, trees, and hash tables, as well as fundamental algorithms like sorting and searching, to highlight their relevance in optimizing microservices functionality [8]. Moreover, Arrays are one of the simplest data structures, providing a fixed-size collection of elements that can be accessed using indices. While they offer efficient data retrieval, their fixed size can be a limitation in dynamic microservices environments where the volume of data may fluctuate. Linked lists, on the other hand, offer more flexibility as they allow for dynamic memory allocation. Each element in a linked list points to the next, facilitating efficient insertions and deletions. In microservices, linked lists can be beneficial for managing collections of data that require frequent updates, as they minimize the need for costly data shifts associated with arrays. Trees, particularly binary trees and their variants (e.g., binary search trees, AVL trees), are essential for hierarchical data representation and enable efficient searching and sorting operations. In a microservices context, trees can be utilized to manage structured data, such as organizational hierarchies or product catalogs, allowing for quick access and retrieval. For scenarios requiring fast lookups, hash tables provide average-case constant time complexity for insertions, deletions, and lookups, making them ideal for scenarios where microservices need to access frequently used data swiftly. Besides, Algorithms also play a crucial role in optimizing microservices performance. Sorting algorithms, such as quicksort and mergesort, help organize data efficiently, while searching algorithms, such as binary search, facilitate quick data retrieval. The implementation of these algorithms can significantly enhance the responsiveness of microservices, ensuring that they can handle user requests in real time. Moreover, understanding the complexities associated with various algorithms allows developers to choose the most suitable options based on the specific needs of their microservices [8-11].

Furthermore, data structures and algorithms are integral to the effective functioning of microservices architectures. By selecting appropriate data structures and implementing efficient algorithms, organizations can optimize data handling, enhance performance, and ensure that microservices operate efficiently. This focus on data efficiency is essential for maintaining a responsive and scalable architecture, ultimately contributing to the overall success of microservices implementations.

The theoretical foundation of microservices architecture is grounded in several key principles that facilitate the development of scalable and maintainable software systems. At its core, microservices architecture embodies the concept of service autonomy, which allows individual services to operate independently while communicating through well-defined APIs [9-10]. This independence is crucial as it empowers development teams to deploy and scale services without affecting the entire application, fostering agility and enhancing fault isolation. Furthermore, microservices rely on decentralized data management, which stands in stark contrast to traditional monolithic systems where a single database is often the bottleneck for performance and scalability. By allowing each microservice to manage its own data, organizations can tailor their data storage solutions to the specific requirements of each service, optimizing resource allocation and promoting efficiency [10].

In addition to these foundational principles, software design models such as Domain-Driven Design (DDD) and Model-View-Controller (MVC) provide a structured approach to creating microservices that are aligned with business objectives. DDD emphasizes the importance of understanding the business domain, encouraging collaboration between technical and non-technical stakeholders to develop services that accurately reflect business needs. This alignment ensures that the microservices not only serve technical functions but also deliver real value to the organization. The MVC design pattern further complements this by promoting a clear separation of concerns, allowing developers to manage application components independently, thus enhancing maintainability and scalability [9-13].

Moreover, the integration of data structures and algorithms is essential for optimizing the performance of microservices. Efficient data handling directly impacts the responsiveness and scalability of services, making it imperative for developers to select appropriate data structures and implement effective algorithms. By leveraging structures such as trees and hash tables, as well as employing algorithms for sorting and searching, microservices can achieve high levels of performance while ensuring quick access to data.

In summary, the theoretical underpinnings of microservices architecture encapsulate principles of autonomy, decentralized data management, and the application of structured design models, all of which contribute to building effective and resilient software systems. This theoretical framework not only supports the technical aspects of microservices but also aligns them with organizational goals, ultimately driving innovation and enhancing business agility [13-15].

III. METHODS AND MATERIALS

This study adopts a qualitative research model, emphasizing an in-depth exploration of the perspectives and experiences of software architects and developers regarding microservices architecture. The qualitative approach is particularly suited to this investigation because it allows for a nuanced understanding of how specific competencies, such as knowledge of software design models and algorithms, influence the effectiveness of microservices implementations. By focusing on the lived experiences of professionals in the field, the research aims to uncover rich, detailed insights that quantitative methods might overlook. This model enables the exploration of complex interactions and relationships within the realm of microservices, highlighting the significance of both technical and contextual factors in shaping architectural decisions and practices. (see **Figure 1**).

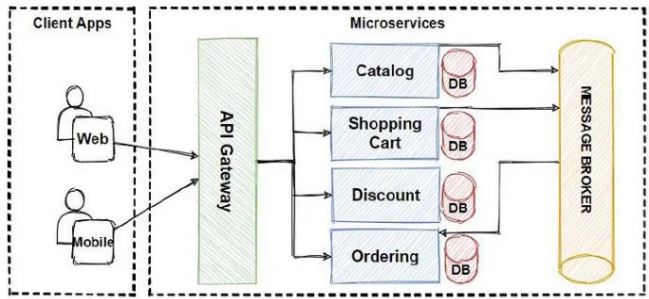

**Figure 1.** Microservices Architecture Design Process.

### Research Design

The research design is structured around semi-structured interviews, allowing participants to share their insights while guiding the conversation toward key themes related to software design and algorithmic proficiency. This flexibility is vital for capturing the diverse experiences and opinions of participants from various industries, ensuring that the study reflects a broad spectrum of knowledge and practices. The semi-structured format facilitates an organic flow of discussion, allowing interviewees to elaborate on their points and share relevant anecdotes. This design choice not only fosters a comfortable environment for participants but also encourages deeper reflection on their experiences with microservices architecture, thereby enriching the data collected.

### Data Collection Sampling

Participants for this study were selected through a purposive sampling technique, targeting software architects and developers with demonstrable experience in microservices architecture. This targeted approach ensures that the insights gathered are relevant and informed by practical experience. The selection criteria focused on professionals who have actively engaged in the design, development, or management of microservices within their organizations. By including individuals from diverse industries, the research captures a variety of perspectives, enhancing the generalizability of the findings. The sample size was determined based on data saturation, ensuring that enough interviews were conducted to identify recurring themes and insights while maintaining a manageable scope for in-depth analysis.

### Data Analysis Process

The data collected from interviews were subjected to thematic analysis, a process that involves identifying, analyzing, and reporting patterns within qualitative data. This approach allows for the systematic organization of data into meaningful themes that reflect the participants' experiences and insights. The analysis began with transcribing the interviews verbatim, followed by coding the data to categorize responses into initial themes. Subsequent rounds of analysis involved refining these themes to capture the nuances of participants' experiences more accurately. By employing a rigorous coding process, the study aimed to ensure reliability and validity in the findings. This thematic analysis not only illuminated the impact of software design models and algorithmic capabilities on microservices architecture but also provided a framework for understanding the broader implications of these insights within the software development landscape [15].

## IV. RESULTS AND DISCUSSION

Microservices architecture is a modern software development approach that structures applications as a collection of loosely coupled, independently deployable services. Each service is designed to perform a specific business function and operates within its own context, which enables organizations to develop, deploy, and scale applications more efficiently. This architectural style contrasts sharply with traditional monolithic architectures, where all components are tightly integrated and deployed as a single unit. In microservices architecture, services communicate through well-defined APIs, promoting a clear separation of concerns that enhances maintainability and flexibility [13-17].

One of the key benefits of microservices architecture is its ability to facilitate scalability. Organizations can scale individual services based on demand without needing to scale the entire application. This flexibility allows teams to allocate resources more effectively and respond quickly to changes in user traffic. Additionally, microservices enable teams to adopt different technologies and frameworks tailored to the specific needs of each service, fostering innovation and improving overall performance.

Moreover, microservices architecture enhances agility in software development. With smaller, focused teams responsible for individual services, organizations can implement changes, deploy updates, and release new features at a much faster pace. This speed is crucial in today's competitive landscape, where businesses must continuously adapt to evolving market conditions and customer expectations [15].

In summary, microservices architecture offers a modular and scalable approach to application development, allowing organizations to build and maintain complex systems with greater ease and efficiency. Its emphasis on service autonomy, flexibility, and responsiveness positions it as a preferred choice for modern software development, enabling businesses to innovate rapidly while delivering high-quality services to their customers.

### Enhanced Scalability

The findings of this study underscore that organizations possessing robust capabilities in software design models significantly enhance their microservices architecture's scalability. Participants consistently emphasized the importance of effective service decomposition, a process facilitated by a comprehensive understanding of design models such as Domain-Driven Design (DDD). By breaking down applications into smaller, focused services, organizations can allocate resources more efficiently and manage workloads effectively [15-17]. This decomposition allows for independent scaling of services, meaning that as demand for specific functionalities increases, organizations can scale those particular services without needing to scale the entire application. Such flexibility is crucial in today's rapidly evolving business landscape, where applications must respond quickly to varying user demands and traffic patterns.

Moreover, scalability in microservices architecture is not solely about handling increased load; it also pertains to the ability to evolve and integrate new functionalities without disrupting existing services. Participants noted that a solid foundation in design principles enables teams to incorporate new services or features seamlessly, which is essential for organizations aiming to innovate continually. This adaptability is a direct result of well-structured service boundaries and clear interfaces, which facilitate communication and integration between services. In essence, the study highlights that organizations that prioritize strong software design models are better positioned to build scalable microservices architectures that can grow alongside their business needs, ensuring long-term sustainability and competitiveness [17-19].

**Improved Performance**

The research findings also reveal that organizations with a strong focus on efficient data structures and algorithms experience notable improvements in the performance of their microservices architectures. Participants shared insights into how optimized data handling directly impacts service responsiveness and overall system latency. By utilizing appropriate data structures, such as hash tables or balanced trees, organizations can enhance data retrieval times and optimize resource usage, which is critical for maintaining high-performance applications. Additionally, implementing effective algorithms for data processing—whether it be through advanced sorting techniques or efficient querying methods—can further minimize delays in service interactions.

Participants highlighted that the choice of data structures and algorithms is crucial during the design phase of microservices. For instance, selecting the right data structure for a specific service can significantly reduce computational overhead and enhance data management. This optimization leads to lower latency, which is particularly important in scenarios where microservices interact frequently, such as in e-commerce platforms or real-time analytics applications. The study emphasizes that organizations prioritizing the development and implementation of efficient data structures and algorithms can achieve faster response times and a more seamless user experience, ultimately contributing to greater customer satisfaction and retention [15-17].

**Greater Flexibility**

Another significant finding of this research is that organizations emphasizing strong software design principles exhibit greater flexibility in adapting their microservices architectures to changing business requirements. Participants noted that a clear understanding of design models enables teams to implement changes swiftly without compromising the integrity of existing services. This flexibility is paramount in a business environment where market conditions, customer needs, and technological advancements can shift rapidly. By adhering to sound design principles, organizations can respond proactively to these changes, ensuring that their microservices can evolve alongside their strategic objectives [16-18].

Furthermore, the ability to adapt microservices easily translates to improved agility within development teams. Participants expressed that organizations that foster a culture of continuous improvement and innovation are more likely to leverage their design knowledge to pivot quickly in response to new challenges or opportunities. This adaptability not only enhances operational efficiency but also positions organizations to capitalize on emerging trends and technologies. The study illustrates that the integration of strong design principles and practices into the development of microservices architectures empowers organizations to maintain a competitive edge, fostering an environment where agility and innovation are at the forefront of their operational strategies. Ultimately, the findings suggest that flexibility, derived from a solid understanding of design models, is a crucial determinant of success in the realm of microservices architecture [19-20].

## V. CONCLUSIONS

In conclusion, the study highlights the critical importance of software design models and data structures and algorithms in the effective implementation of microservices architecture. A strong understanding of design models facilitates service decomposition, enabling organizations to create scalable and maintainable systems [16]. This capability allows businesses to allocate resources more efficiently and respond dynamically to changing demands. Furthermore, the strategic use of optimized data structures and algorithms significantly enhances performance by improving data processing speeds and reducing latency in service interactions. Together, these competencies empower organizations to innovate rapidly and maintain competitive advantages in fast-paced markets. As companies increasingly adopt microservices architecture, prioritizing training in design principles and algorithmic efficiency will be essential for achieving operational excellence and driving future growth. The findings underscore the necessity for software teams to integrate these foundational elements into their development practices, ensuring robust, agile, and high-performing microservices solutions. Overall, this research contributes to a deeper understanding of how these theoretical frameworks translate into practical advantages in software development and architecture [17-22].

**Findings Implications**

The findings of this study indicate that organizations with strong software design model capabilities and a solid grasp of data structures and algorithms experience significant advantages in microservices architecture design [20]. Specifically, the ability to effectively decompose services leads to enhanced scalability, allowing businesses to allocate resources efficiently based on demand. Moreover, efficient data handling and processing improve performance, reducing latency and ensuring faster service interactions. These implications highlight the importance of investing in training and development for software architects and developers to deepen their understanding of design principles and algorithmic efficiency. As organizations increasingly adopt microservices architecture, the need for these competencies becomes essential for achieving operational excellence and maintaining a competitive edge in rapidly changing markets.

This research contributes to the existing body of knowledge by providing empirical evidence on the critical role of software design models and data structures in microservices architecture. By conducting interviews with industry professionals, the study illuminates the real-world applications of theoretical concepts, bridging the gap between academia and practice. Furthermore, the findings offer practical insights for organizations looking to enhance their microservices strategies [18-21]. The identification of specific design principles and algorithmic techniques that drive scalability and performance provides a valuable resource for practitioners aiming to optimize their development processes. Overall, this research not only enriches academic discourse but also serves as a guide for practitioners seeking to implement effective microservices architectures.

While this study offers valuable insights, it is essential to acknowledge its limitations. The qualitative nature of the research, while providing depth and context, may not capture the full range of experiences and practices across all industries. The sample size, although sufficient for thematic saturation, is limited to participants who have experience with microservices, potentially introducing selection bias. Additionally, the study focuses primarily on software design models and algorithms, potentially overlooking other critical factors influencing microservices architecture, such as organizational culture or team dynamics. Future research could address these limitations by incorporating a more diverse sample and exploring additional variables that impact the effectiveness of microservices implementations [20-22].

**Future Work Directions**

Future research should build upon the findings of this study by exploring the intersection of microservices architecture with other emerging technologies, such as cloud computing and artificial intelligence. Investigating how these technologies can enhance the capabilities of microservices could provide further insights into optimizing application performance and scalability. Additionally, longitudinal studies examining the long-term effects of adopting microservices architecture in various organizational contexts would contribute to a deeper understanding of its impacts over time. Moreover, research could explore the relationship between organizational culture and the successful implementation of microservices, identifying best practices for fostering an environment conducive to innovation and agility. By addressing these areas, future work can provide comprehensive insights that further advance the field of microservices architecture [21-22].

ACKNOWLEDGMENT

This research has been supported/partially supported Solbridge International School of Business, Woosong university, Thanks to all contributors.

ORCID

Jun Cui  https://orcid.org/0009-0002-9693-9145